\input harvmac
\noblackbox
\font\ticp=cmcsc10
 
\def\Title#1#2{\rightline{#1}\ifx\answ\bigans\nopagenumbers\pageno0\vskip1in
\else\pageno1\vskip.8in\fi \centerline{\titlefont #2}\vskip .5in}

\font\ticp=cmcsc10
\font\ttsmall=cmtt10 at 8pt

%
%
\def\[{\left [}
\def\]{\right ]}
\def\({\left (}
\def\){\right )}
\def\p{\partial}
\def\a{\rightarrow}
\def\t{\tilde}
\def\A{$AdS_2$}

\lref\cvla{M. Cvetic and F. Larsen, hep-th/9805097, Nucl. Phys.
B531 (1998) 239; hep-th/9805146, Phys. Rev. Lett. 82 (1999) 484.}
\lref\gmt{J. P. Gauntlett, R. C. Myers, and P. K. Townsend, 
hep-th/9809065, Phys. Rev. D59 (1999) 025001; hep-th/9810204, 
Class. Quant. Grav. 16 (1999) 1.}
\lref\krw{R. Kallosh, A. Rajaraman and W. K. Wong, hep-th/9611094,
Phys. Rev. D55 (1997) 3246.}
\lref\hht{S. W. Hawking, C. J. Hunter, and M. M. Taylor-Robinson,
 hep-th/9811056,
Phys. Rev. D59 (1999) 064005.}
\lref\mms{J. Maldacena, J. Michelson, and A. Strominger, hep-th/9812073,
JHEP 9902 (1999) 011.}
\lref\wald{R. Wald, {\it General Relativity}, U. of Chicago Press (1984) p.328.}
\lref\hopo{G. Horowitz and J. Polchinski, hep-th/9712146, Phys. Rev. D55 (1997)
6189.}
\lref\sing{S. Hawking and R. Penrose, Proc. Roy. Soc. Lond. A314 (1970) 529.}
\lref\det{S. Detweiler, Astrophys. J. 239 (1980) 292; 
S. Detweiler and R. Ove, Phys. Rev. Lett. 51 (1983) 67. }
\lref\rusu{J. Russo and L. Susskind, hep-th/9405117, Nucl. Phys. B437 
(1995) 611.}
\lref\mype{R. Myers and M. Perry, Ann. Phys. 172 (1986) 304.}
\lref\juan{J. Maldacena, hep-th/9711200, Adv. Theor. Math. Phys. 2 (1998) 231.}
\lref\holo{G. 't Hooft, ``Dimensional Reduction in Quantum Gravity",
gr-qc/9310026; L Susskind, hep-th/9409089,
J. Math. Phys. 36 (1995) 6377.}
\lref\joe{J. Polchinski, ``S-Matrices from AdS Spacetime", hep-th/9901076; 
L. Susskind, ``Holography in the Flat Space Limit", hep-th/9901079.}
\lref\adst{A. Strominger, hep-th/9809027, JHEP 9901 (1999) 007;
M Hotta, ``Asymptotic Isometry and Two Dimensional Anti-de Sitter Gravity",
gr-qc/9809035;
T. Nakatsu and N. Yokoi, hep-th/9812047, Mod. Phys. Lett. A14 (1999) 147;
  G. Gibbons and P. Townsend, ``Black Holes and Calogero Models",
hep-th/9812034; M. Cadoni and S. Mignemi, ``Asymptotic symmetries of $AdS_2$
 and conformal group in d=1", hep-th/9902040; M Spradlin and
A. Strominger, ``Vacuum States for $AdS_2$ Black Holes", hep-th/9904143.}
\lref\page{D. Page, Phys. Rev. D14 (1976) 3260.}
\lref\carter{B. Carter, Commun. Math. Phys. 10 (1968) 280; C. Misner,
K. Thorne, and J. Wheeler, {\it Gravitation}, Freeman (1973).}
\lref\bawa{J. Bardeen and R. Wagoner, Astrophys. J. 167 (1971) 359.}
\lref\bardeen{J. Bardeen, in {\it Black Holes}, C. DeWitt and B. DeWitt eds.,
Gordon and Breach (1973).}
\lref\hlm{G. Horowitz, D. Lowe and J. Maldacena, hep-th/9603195, 
Phys. Rev. Lett. 77 (1996) 430.}

%
\baselineskip 16pt
\Title{\vbox{\baselineskip12pt
\line{\hfil   NSF-ITP-99-29}
\line{\hfil \tt hep-th/9905099} }}
{\vbox{
{\centerline{The Extreme Kerr Throat Geometry:}
\vskip .5cm
\centerline{A Vacuum Analog of $AdS_2\times S^2$}}
}}
\centerline{\ticp James Bardeen and Gary T. Horowitz\footnote{}{\ttsmall
jbardeen@itp.ucsb.edu, gary@cosmic.physics.ucsb.edu}}
\bigskip
\vskip.1in
\centerline{\it Institute for Theoretical Physics, University of California,
Santa Barbara, CA 93106, USA}
\bigskip
\centerline{\bf Abstract}
\bigskip
We study the near horizon limit of a four dimensional extreme rotating
black hole.
The limiting metric is a  completely nonsingular vacuum solution, with an
enhanced symmetry group $SL(2,R)\times U(1)$. We show that many of the
properties of this solution are similar to the $AdS_2\times S^2$ geometry
 arising in the near horizon limit of extreme 
charged black holes. In particular, the boundary at infinity is a timelike
surface. This suggests the possibility of a dual quantum mechanical description.
 A five dimensional generalization is also discussed.

\Date{May, 1999}

\newsec{Introduction}

There is growing evidence in support of the conjecture that string theory
with asymptotically Anti de Sitter (AdS) boundary conditions is completely
described by a dual 
conformal field theory (CFT) \juan. One of the features of AdS that makes this
correspondence possible is that the (conformal) boundary at infinity is a 
timelike surface. In many applications, one can imagine the dual CFT living
on this timelike boundary. For this reason, the AdS/CFT correspondence is
often called holographic \holo. In contrast, asymptotically flat boundary
conditions lead to a boundary at infinity consisting of two null surfaces
(together with a point at spatial infinity). It is far from clear what form
a holographic description will take in this case\foot{For a discussion of
obtaining this as a limit of the AdS/CFT correspondence see 
\joe.}.

 We describe below some vacuum spacetimes
with asymptotic structure similar to AdS. They are
obtained by taking the near horizon geometry of extreme rotating
black hole solutions.  In four dimensions, there is a one parameter family
 of solutions
labeled by the total angular momentum $J$. These vacuum spacetimes
are completely nonsingular, and 
can be used to construct classical solutions
to all string theories. One simply takes the product with a Ricci flat
internal space to obtain a ten (or eleven)
dimensional solution. Since the curvature
is bounded everywhere, and small for large $J$,
the $\alpha'$ (or M-theory) 
corrections should be small. One advantage over the AdS solutions
is that there are no background
Ramond-Ramond fields, so string propagation in these backgrounds
 is straightforward. However an obvious disadvantage is that these solutions
do not admit covariantly constant spinors, and hence are not supersymmetric.
The effect of quantum corrections remains to be investigated.

It is tempting to speculate that there is some type of dual CFT description
 of string theory on spacetimes which
 asymptotically approach these vacuum solutions. 
Starting with the four dimensional Kerr solution,
one obtains a vacuum solution which resembles $AdS_2\times S^2$. It has
a symmetry group $SL(2,R) \times U(1)$ and a timelike boundary at infinity.
Since
the boundary of \A\ is one dimensional, one expects the dual theory to have
a conformal quantum mechanical description. Unfortunately, despite much effort
\refs{\adst,\mms}
 the $AdS_2/CFT_1$ correspondence is still poorly understood. We
will not be able to describe the dual theory in our case, although we
will make some comments in section five. 

Since the AdS/CFT duality is better understood in higher dimensions,
it is natural to ask whether spacetimes analogous to higher dimensional
AdS arise in the near horizon limits of rotating black holes in higher 
dimensions. 
Several people have studied the near horizon geometry of rotating charged
black holes and p-branes \refs{\cvla,\gmt,\krw,\hht}. However, in all these
cases, the charge plays an essential role, and most solutions are asymptotically
AdS. What about higher dimensional analogs of the Kerr solution \mype?
If the spacetime
has only one component of angular momentum nonzero, then there is no
extremal black hole in more than five dimensions. A given mass
black hole can have arbitrarily large angular momentum. If all components
of the angular momentum
are nonzero, then there is an extremal limit, but we expect the near horizon 
geometry will still resemble \A.  This is because the effective cosmological
constant in these vacuum solutions can be thought of as arising
 from the off diagonal terms in a
Kaluza-Klein reduction of the metric. This always produces a two form Maxwell
field which can act like a cosmological constant in two dimensions only. To
obtain $AdS_n$ for $n>2$ one would need a higher rank form which does not
arise naturally in a vacuum solution.

In the next section we derive the near horizon limit of the extreme Kerr 
solution and show that it has enhanced symmetry.
We also discuss geodesics, and find that timelike geodesics 
with sufficiently large angular momentum can escape to infinity. 
However, just like radial null geodesics in AdS, they do so in finite
coordinate time but infinite affine parameter. This shows that the throat
solution is geodesically complete, and has a timelike boundary.

The possibility of viewing the vicinity of the extreme Kerr horizon
as a complete vacuum spacetime in its own right was suggested by the
results of Bardeen and Wagoner \bawa\ (see also \bardeen). They studied
the exterior metric of a uniformly rotating disk in the extreme relativistic
limit. If the relativistic limit (infinite redshift from the center of the
disk) is taken before taking the limit of infinite affine distance from the
 disk, the asymptotic geometry is the throat of an extreme Kerr black hole
rather than an asymptotically flat spacetime. 

In section three, we discuss propagation of a massless scalar test field
in this background. Most modes have discrete frequencies and are confined.
However, a few modes with large azimuthal angular momentum can propagate
to infinity. We will argue that this can lead to an analog of superradiance
where a wave can scatter and return with more energy than it started with.
When backreaction is included, the throat
 solution is definitely unstable. We give
a general argument (independent of superradiance)
that nearby solutions will be singular. This argument also
shows that $AdS_2\times S^2$ is unstable, but not higher
 dimensional AdS spacetimes. In \mms\ another argument for the 
instability of $AdS_2\times S^2$ is given, and it is suggested
that the dual theory may describe only ``ground states" of
string theory, and not finite energy excitations. The same may be true
in our case as well. Alternatively, the instability may simply indicate
that the addition of any amount of energy produces a black hole. 
The dual theory could perhaps describe these black states as well.

Our results can be extended to the general Kerr Newman solution
describing a charged
rotating black hole. The near horizon limit of this solution is briefly 
discussed in section four. In the following section we make some
comments about the dual quantum mechanical theory. Perhaps the key observation
is that the area of the event horizon is related to the effective
cosmological constant of \A\ in a universal way that is independent of
whether the extreme black hole has only angular momentum, charge or both.
This suggests that the quantum mechanical theories which are dual to these
backgrounds are closely related.
 
Finally, in section six, we discuss the near horizon limit of the 
five dimensional extreme Kerr solution. The resulting geometry
is more complicated, but
qualitatively similar to the
four dimensional case. The limiting solution resembles
$AdS_2\times S^3$.

\newsec{Kerr throat solution and its properties}

We begin with the (four dimensional)
Kerr metric in Boyer-Linquist coordinates:
\eqn\kerr{
ds^2 = -e^{2\nu} d\t t^2 + e^{2\psi}(d\t \phi -  \omega d\t t)^2 +
\rho^2 (\Delta^{-1} d\t r^2 + d\theta^2) }
where
\eqn\deff{\rho^2 \equiv \t r^2 +a^2\cos^2\theta, \qquad
         \Delta \equiv \t r^2-2M\t r +a^2}
and
\eqn\comp{e^{2\nu} = {\Delta \rho^2 \over (\t r^2 + a^2)^2 
  - \Delta a^2 \sin^2\theta}, \qquad
  e^{2\psi} = \Delta \sin^2\theta\  e^{-2\nu}, \qquad
      \omega = {2M\t r a\over \Delta \rho^2} e^{2\nu}}  
The total mass is $M$, and the angular momentum is $J=Ma$ which we will assume
is positive\foot{We have set $G=1$ where $G$ is the four dimensional
 Newton's constant.}. 
The extremal
limit corresponds to $a^2=M^2$, so $\Delta = (\t r-M)^2$ and the event horizon
 is at $\t r=M$.
 The area of the extremal horizon is 
\eqn\area{ A = 8\pi M^2 = 8\pi J}
The value of $ \omega$ at the horizon is called the angular velocity of the
horizon and, in the extremal case, is simply $ \omega = 1/2M$.
Since $g_{\tilde r \tilde r} = \rho^2/\Delta$ it is clear that the
spatial distance to the extremal horizon in a constant $\tilde t$ surface
is infinite. In analogy to the extreme charged black holes, we wish to 
extract the limiting geometry as one moves down this throat. 

To describe this near horizon geometry,
we set
\eqn\limit{ \t r = M + \lambda r ,\qquad  \t t = {t\over\lambda},
\qquad \t \phi = \phi + {t\over 2M\lambda}}
and take the limit $\lambda \a 0$. The shift from $\t \phi$ to $\phi$
makes
$\p/\p t$ tangent to the horizon. In other words, the coordinates
corotate with the horizon. The result is
\eqn\throat{\eqalign{ ds^2 =&\({1+ \cos^2{\theta}\over 2}\) 
\[ -{r^2\over r_0^2} dt^2 +
  {r_0^2 \over  r^2} dr^2 + r_0^2 d\theta^2\] \cr
& + {2 r_0^2 \sin^2\theta \over 1+\cos^2 \theta}
	\( d\phi + {r\over r_0^2} dt\)^2 \cr }}
where we have defined $r_0^2 \equiv 2M^2$. This spacetime is no longer asymptotically
flat. We will show that it is similar to $AdS_2\times S^2$ in many respects.
For example, if one sets $\theta=0$ (or $\pi$) one sees
that the spacetime along the axis is {\it precisely} $AdS_2$.
It  is clear that by rescaling $t$, one
can ensure that $r_0$ only appears as an overall factor in front of the metric.

It is easy to see that \throat\ has enhanced symmetry. In addition to the
$\p/\p t$ and $\p/\p \phi$ symmetries present in Kerr, \throat\
is clearly invariant under $r\a cr, \ t \a t/c$ for any constant $c$. 
So \throat\ has the dilation symmetry 
of $AdS_2$. It is less obvious, but still true, that \throat\ is also invariant
under (an analog of) the global time translation in $AdS_2$. To see this,
note that the 
$(r,t)$ coordinates are analogous to 
Poincare coordinates on $AdS_2$.
To find the extra global time translation symmetry, we introduce new
coordinates which are related to $(r,t)$ in the same way that the global
coordinates of $AdS_2$ are related to the Poincare coordinates.  
For simplicity, we set $r_0 =1$. Let
\eqn\newcoor{ r = (1+y^2)^{1/2}\cos\tau +y, \qquad t = {(1+y^2)^{1/2}\sin\tau
\over r}.}
The new axial angle coordinate $\varphi$ is chosen so that $g_{\varphi y} = 0$,
with the result
\eqn\phitran{ \phi = \varphi +\log \Big|{\cos\tau +y\sin\tau \over {1 +
(1 +y^2)^{1/2}\sin\tau}}\Big|.}
In these new coordinates, the throat metric \throat\
 takes the form 
\eqn\glob{\eqalign{ds^2 =& \({1+ \cos^2{\theta}\over 2}\) \[-(1+y^2) d\tau^2 +
  {dy^2 \over  1+y^2}  +  d\theta^2\] \cr
&+ {2 \sin^2\theta \over 1+\cos^2 \theta}( d\varphi + y d\tau)^2 .\cr}}
Note that the $\tau = 0$ hypersurface coincides with a $t = 0$
hypersurface, and that $\varphi = \phi$ on this hypersurface (and also
at infinity, for all time).

The throat solution \glob\ thus 
has all of the symmetries of \A\ plus translations
in $\varphi$: Its isometry group is $SL(2,R) \times U(1)$. All geometric
quantities depend only on $\theta$.
We will show below that the coordinates in \glob\
(with $-\infty < \tau< \infty $,
$-\infty <y< \infty$) cover the entire spacetime.
The surfaces of constant $\tau$ are always
spacelike, so $\tau$ is a global time function and this spacetime has no closed
timelike curves. However
the Killing field $\p/\p \tau$ is not timelike everywhere.
It is timelike
for all $\theta$ when $y^2 < 1/3$, but asymptotically is spacelike for
$\sin\theta > (1 +\cos^2\theta)/2$, or $\sin\theta > 0.536$, within
32.4 degrees of the equatorial plane. 
This is a consequence of the rotation and is
analogous to the ergosphere in the extreme Kerr solution\foot{This is not
exactly the same as the original ergosphere, since $\p/\p \tau$ is
not a Killing field in the full Kerr metric.}.  
It will be convenient to define the vector field
\eqn\defchi{\chi = {\p\over \p \tau} - y {\p\over \p \varphi}.}
$\chi$
 is a future directed timelike vector everywhere, which is orthogonal to 
the surfaces of constant $\tau$ and tangent to the surfaces of constant $y$.

There are two boundaries at infinity corresponding to $y=\pm \infty$.
Since a surface of constant $y$ is
always timelike, the limiting surfaces $y=\pm \infty$ must be timelike or
null. The boundary will be timelike if geodesics can reach it in a finite
value of $\tau$. We now show that this is the case. In the process, we will 
also show that \glob\ is geodesically complete. Hence the solution is 
nonsingular, and the coordinates
$(\tau,y,\theta,\varphi)$ cover the entire spacetime.

Since we are mostly
interested in the asymptotic properties of the geodesics it
suffices to consider geodesics with constant $\theta$. This
corresponds to $\theta =0,\pi/2$.
It is clear that test particles moving along the axis $\theta=0$
will behave exactly as in $AdS_2$. In particular, timelike geodesics never
reach infinity, and null geodesics reach infinity in a finite time.
The equatorial plane, $\theta =\pi/2$, is a three dimensional homogeneous space
with symmetry group $SL(2,R) \times U(1)$. It is a twisted product
of $AdS_2$ and a circle of constant radius.
Consider a geodesic with momentum
$P = \dot \tau (\p/\p \tau) + \dot y (\p/\p y) +
\dot \varphi (\p/\p \varphi)$, where a dot denotes derivative with respect to
an affine parameter. The conserved quantities are
\eqn\conserve{
L =P \cdot \p/\p \varphi = 2(\dot \varphi + y \dot \tau), \qquad
E = -P \cdot \p/\p \tau = {1+y^2\over 2} \dot\tau  - Ly }
Setting $g_{\mu\nu} P^\mu P^\nu = -\mu^2$ yields
\eqn\radial{
\dot y^2 -4(E+Ly)^2 +(2\mu^2 +L^2)(1+y^2)=0  }
Geodesics with zero angular momentum again behave exactly like geodesics in \A.
That is,  massive particles feel an infinite potential barrier and stay
confined to finite $|y|$. Massless particles can reach infinity, but 
since $y$ is proportional to the affine parameter, these geodesics are
obviously complete. One can easily verify that these geodesics reach 
infinity in a finite time $\tau$. Timelike geodesics with $L^2 < 2\mu^2/3$
are also confined.  However, geodesics with $L^2 > 2\mu^2/3$ can escape to 
infinity. This is a qualitatively new feature of \glob\ which is
 not present in $AdS_2\times
S^2$. Since $\chi$ \defchi\ is a future directed timelike vector everywhere,
it follows that $-P\cdot \chi >0$ which implies $E+yL>0$. Thus, a geodesic
with $L>0$ can only escape to $y=+\infty$, while a geodesic with $L<0$
can only escape to $y=-\infty$. 
Since $\dot y \propto y$ asymptotically these geodesics are also complete,
and reach infinity in a finite value of $\tau$. Geodesics with
$L^2 = 2\mu^2/3$ satisfy $\dot y^2 = 8ELy$ asymptotically. So once again
if $L>0$,
these geodesics can reach $y=\infty$ but not $y=-\infty$. For $L<0$,
the situation is reversed. These geodesics are also complete and reach
infinity in finite $\tau$.

For product spacetimes such as $AdS_2\times S^2$, one can remove the
angular directions, and conformally rescale \A\ to view infinity as a finite
boundary. Then the Killing fields of \A\ become conformal symmetries of the
boundary. One cannot do this for the throat solution since the geometry
does not approach a product metric asymptotically. If one wants to bring 
infinity in to a finite distance, the best one can do
is to rescale the entire metric, although the conformal metric is no longer
smooth at the boundary. For convenience, we start with the
metric in Poincare coordinates \throat\ (although we set $r_0=1$). 
Multiplying by $1/r^2$ and setting $x=1/r$ the
metric becomes
$$ds^2 = \({1+ \cos^2{\theta}\over 2}\) \[- d t^2 +
  dx^2   + x^2 d\theta^2\] $$
\eqn\conf{ 
+ {2 \sin^2\theta \over 1+\cos^2 \theta}(x d\phi +  d t)^2 .}
Despite its simple appearence, this conformal metric has a curvature
singularity at $x=0$. The four Killing fields $\xi_i^\mu$ of \throat\ 
are
$$\xi_1 = {\p\over \p \phi}, \qquad \xi_2 = {\p\over \p t},
 \qquad \xi_3 = t{\p\over \p t} -r {\p\over \p r}$$
\eqn\kf{\xi_4 = \({1\over 2r^2} + {t^2\over 2} \) {\p\over \p t}
-tr {\p\over \p r} -{1\over r} {\p\over \p \phi}  }
Since they are Killing fields of \throat, they are 
conformal Killing fields of \conf\ with action
${\cal L}_{\xi} (r^{-2} g_{\mu\nu}) = ({\cal L}_{\xi} r^{-2}) g_{\mu\nu} =
(-2 \xi^r /r)(r^{-2} g_{\mu\nu})$. The net result is that $\xi_1$ and
$\xi_2$ remain Killing fields of the rescaled metric, $\xi_3$ multiplies
the metric \conf\ by two, and $\xi_4$ multiplies the metric by $2t$.
 Thus the last
three generate the conformal group on a line with metric $-dt^2$. 
This can be realized explicitly by introducing a cutoff at $x=\epsilon$
and shifting $\phi = \hat \phi -t/ \epsilon$. Then the metric on the 
$x=\epsilon$ surface is essentially the product of a line and a very
small sphere.

\newsec{Modes of a Massless Scalar Field}

An important feature of test fields on $AdS_2\times S^2$ is that they
naturally obey reflecting boundary conditions at infinity. (This is true for
all modes with nontrivial dependence on $S^2$.)  The wave
solutions have a discrete spectrum, with no oscillatory
behavior near infinity.
 We will show that a similar confinement holds for axisymmetric 
modes in the
extremal Kerr throat geometry. However, as might be suspected from the behavior
of geodesics with large angular momentum found in Section 2, some
non-axisymmetric modes do propagate all the way to infinity and
transport energy and angular momentum there. These scattering states have
a continuous spectrum. This behavior is
expected be generic for waves of all spins.  We illustrate it with a
discussion of solutions of the massless scalar wave equation.

The scalar wave equation for a spacetime described by a metric of the form
\kerr\ can be written as 
$${1 \over \rho^2}{\p \over \p \t r}\[{\Delta {\p \Psi \over \p \t r} }\] +
{1 \over \rho^2 \sin\theta} {\p \over \p \theta} \[{\sin\theta {\p \Psi \over
\p \theta} }\]$$
\eqn\scgen{ + \[e^{-2\nu} (\t \sigma -m  \omega)^2 -m^2 e^{-2\psi} \] \Psi
= 0}
assuming harmonic time and axial angle dependence of the form
\eqn\harmonic{\Psi = \Psi (\t r,\theta)e^{im\t \phi -i\t \sigma \t t}.}
In taking the near horizon limit \limit,
 the throat frequency $\sigma$ is related
to the Kerr frequency $\t \sigma$ by
\eqn\freqtr{\t \sigma -m\Omega_H =\lambda\sigma.}
where $\Omega_H$ is the angular velocity of the horizon.
Therefore, as $\lambda \a 0$ all finite frequencies in the throat correspond to
the single frequency $\t \sigma = m\Omega_H = m/2M$ in the exterior.

Modes in Kerr with $\t \sigma = m\Omega_H $ are special since they 
are totally reflected, and not absorbed by the black hole. This can be seen 
from the following argument, which applies to all Kerr black holes, and not just
extremal ones (see \wald).
Let $\xi^\mu = (\p /\p \t t)^\mu$ be the usual
stationary Killing field. The null vector tangent to the horizon of a Kerr
black hole is $l^\mu = \xi^\mu + \Omega_H (\p/ \p \t\phi)^\mu$.
Since $l^\mu \xi_\mu =0$ on the horizon,
 the energy flux entering the black hole is obtained by integrating
$T_{\mu\nu}\xi^\mu l^\nu = (\xi^\mu \p_\mu \Psi)(l^\nu \p_\nu \Psi)$ over the 
horizon. This is proportional to $\t \sigma(\t \sigma -m\Omega_H)$. So modes
with $\t \sigma > m\Omega_H$ have a positive energy flux into the black hole
and correspond to normal scattering. Modes with $\t \sigma < m\Omega_H$
have a negative energy flux and correspond to superradiant scattering: The
outgoing wave has more energy than the incoming one. Modes with exactly
$\t \sigma = m\Omega_H$ are on the borderline. No energy is absorbed by
the black hole. The modes we will study in the throat geometry can be 
thought of as obtained by starting with a mode in the extreme Kerr metric
 satisfying
$\t \sigma =m\Omega_H +\lambda\sigma$. For small $\lambda$, most of this
wave is reflected and only a small part remains near the horizon. After
the rescaling \limit, the wave near the horizon remains nonzero
in the limit $\lambda
\a 0$, and completely decouples from the wave in the asymptotically flat region.

The form of the throat metric in Poincare and global $AdS_2$ coordinates becomes
 identical
 as the respective
radial coordinates $r$ and $y$ become large, so the asymptotic
properties of the wave solutions are the same in both coordinate systems.
For definiteness,
we will use the global coordinates. Since the metric \glob\ is of the general
form \kerr, the wave equation in this background takes the form \scgen\
for suitable choice of metric components.
This wave equation, after multiplication by $\rho^2$, is completely
separable, just as it is for the original Kerr metric.  
Setting $\Psi = Y(y) \Theta(\theta)$ splits the partial
differential equation into the two ordinary differential equations
\eqn\scrad{ {d \over d y} \[ (1 +y^2) {d Y \over d y} \] +\[ {(\sigma +
m y)^2 \over 1 +y^2} +m^2 -K \] Y = 0}
and
\eqn\scang{ {1 \over \sin\theta} {d \over d \theta} \[ \sin\theta
{d \Theta \over d \theta}\] + \[ K - {m^2 \over \sin^2\theta} - {1 \over 4}
m^2 \sin^2\theta \] \Theta = 0.}
The separation constant $K$ has been defined so that the angular equation is
identical to the spheroidal harmonic angular equation familiar from solutions
on the Kerr background, with the Kerr frequency set to the unique value
implied by \freqtr.

The eigensolutions of \scang\ with boundary conditions of
regularity at $\theta =0$ and $\theta =\pi$ are the usual
Legendre functions for $m = 0$, so $K = l(l+1)$.
 But even for non-zero $m$, the solutions are fairly
close to associated Legendre functions, since
${1 \over 4}m^2\sin^2\theta$ is rather small compared to $l(l+1)$ for
$ l\ge |m|$.  
Numerical calculations of the eigenvalues $K$ for a few of the
lowest modes are listed below. We also include the value of 
 $K_{crit} \equiv 2m^2 -1/4$ which we will see marks the dividing line
between the discrete and continuous part of the spectrum.
$$\vbox{\settabs\+  &m  \qquad    & l = m   \qquad      &l = m+1  
  \qquad     &l = m+2 \qquad & l=m+3 \cr
     \+    & m       & l=m         &l=m+1     & l=m+2  & $K_{crit}$  \cr
      \+   & 1       &2.200        &6.143     & 12.133   &  1.75 \cr        
      \+    & 2      &6.855        &12.664    & 20.597   & 7.75 \cr
      \+    &3      &13.995         &21.629    &  31.459  & 17.75 \cr} $$
\centerline{Table 1. Some eigenvalues $K$ of the angular equation \scang.} 
In general,
\eqn\eigval{ K_{l m} \approx l(l+1) + c\ m^2 }
where $c$ is roughly in the range $.13$ to $.22$
and  $l\ge |m|$ are integers.

The radial equation for axisymmetric modes is identical to that for
$AdS_2$.  The natural boundary condition at infinity is $\Psi = 0$.
Then $\Psi \sim |y|^{-l-1}$, and normal modes satisfying the boundary
conditions at both $y = +\infty$ and $y = -\infty$ exist for
discrete real frequencies $\sigma_{l n}$.  Waves are reflected
before they reach infinity.

The non-axisymmetric modes are more interesting.  The asymptotic
solutions for $|y| \gg 1$ are power laws $Y \sim |y|^\alpha$, with
\eqn\powexp{ \alpha = -{1 \over 2} \pm (K -2m^2 +{1 \over 4})^{1/2}.}
For $K > K_{crit}\equiv 2m^2 - 1/4$ both exponents are real, qualitatively like
the axisymmetric modes. Only the more
rapidly decaying solution satisfies the boundary condition, and
global normal modes exist for discrete frequencies.
On the other
hand, if $K < 2m^2 - 1/4$ the exponents are complex, and the
asymptotic solutions are traveling waves.  Positive frequency
modes are outgoing in phase velocity
if Im $\alpha > 0$ and ingoing if Im $\alpha < 0$.
From the above table, it is clear that for $m=1$,
 all eigenvalues $K$ are larger than $K_{crit} = 2m^2- 1/4$, but
for all $|m| > 1$ at least the
$l = |m|$ eigenvalue is less than $K_{crit}$.  This is qualitatively
different from $AdS_2$, since in the absence of rotation
the condition for confinement is $K = l(l+1) > m^2 - 1/4$, which
is always satisfied.

In a WKB approximation to the traveling waves,
 the effective wavenumber $k=-{i\over Y}{dY\over dy}$ is
\eqn\waveno{k = \pm {1\over(1+ y^2)^{1/2}}
\[{(\sigma +m y)^2\over 1+y^2} +m^2 -K\]^{1/2}.}
The group velocity is then
\eqn\group{{d \sigma \over d k} = \pm {(1+y^2)^{3/2}\over \sigma +m y} 
\[{(\sigma +m y)^2\over 1+y^2} +m^2 -K\]^{1/2}.}
The phase velocity is just $\sigma/k$. So
for $m>0$, the phase velocity and group velocity have opposite signs
for large negative $y$. For $m<0$, they have opposite signs for large
positive $y$. This has an important consequence, which we now explain.
 
Since $K$ is always larger than $m^2$, the expression inside the
brackets in \waveno\ always changes sign around $\sigma + my =0$.
This means that the wave encounters a potential barrier and the
WKB approximation breaks down. 
An initial wave with $\sigma > 0$ and $m > 0$ moving in the negative
$y$ direction will be partly reflected and partly transmitted through this
barrier. We now have to discuss the physical boundary conditions on these
waves. First note that all modes vanish at infinity.
Even when $\alpha$ is complex, we still have $Y \sim |y|^{-1/2}$.
But since the volume element on a surface of constant $y$ is proportional
to $|y|$ asymptotically,
 there can still be a nonzero flux of energy and angular momentum to infinity.
We demand that the transmitted wave should be purely outgoing, where
``outgoing" is defined with respect to the physical group 
velocity.

To compute the flux of energy and angular momentum, note that
each Killing vector field of the spacetime produces a conserved flux
4-vector when
contracted with the energy-momentum tensor of the wave.  The axial
Killing field
 is the unique Killing field with closed orbits, and gives the angular momentum
flux vector $J^\mu = T_\varphi^\mu$.  The average rate (per unit global
time $\tau$) of angular momentum transport in the $+y$ direction across
a constant-$y$ surface is equal to
\eqn\angmomfl{\int d\theta d\varphi\
(-g)^{1/2}\ T_\varphi^y = \int d\theta d\varphi 
\sin\theta \ (1 +y^2) {\p \Psi \over \p \varphi}{\p \Psi \over \p y}.}
After averaging over time, and taking the limit $y \a \infty$ with
asymptotically $Y = A_+ y^{i(2m^2 -1/4 - K)^{1/2} - 1/2}$ for the
outgoing positive frequency wave solution,
the angular momentum transport rate becomes
$m(2m^2 -1/4 -K)^{1/2} |A_+|^2$, assuming
the angular harmonics are normalized.  Using the Killing field
$\p /\p\tau$ to define a conserved energy flux, the asymptotic
energy transport rate is $\sigma / m$ times the angular momentum transport
rate calculated above, and for positive frequency waves is outward
for waves with an outward phase velocity.

We now come to the key point. Consider a wave with $\sigma>0,m>0$ which 
starts at large $y$ moving in the negative $y$ direction. After scattering
off the potential barrier, there will be a reflected wave and transmitted wave.
In order for the
transmitted wave to have outgoing group velocity, it must have incoming
phase velocity. Thus there is a incoming flux of energy from $y=-\infty$.
Since this energy is conserved, there must be a corresponding outgoing
flux of energy at $y=\infty$. Since we started with an incoming wave at
$y=\infty$, the only way this is possible is if the reflected wave 
carries more energy than the initial wave.
In this sense, waves in the throat geometry exhibit superradiance.

Another way to understand
the difference in sign between the group velocity and
phase velocity is to consider a local observer
who is at rest with respect to the constant $\tau$ surfaces, and hence has
a four velocity proportional to $\chi$ \defchi. Such a 
zero angular momentum observer (ZAMO) at constant $y = y_o$ and $\theta$,
would assign a frequency to the wave
\eqn\zamofr{ \sigma_{ZAMO} = {\sigma +m y_o \over
[(1 +y_o^2) (1 +\cos^2\theta )/2)]^{1/2}}.}
At large positive $y_o$ with negative $m$ or large negative $y_o$ with
positive $m$ the ZAMO would assign a negative frequency to the wave,
and according to him the direction of the phase velocity 
is the same as the group velocity.   The local
timelike Killing field tangent to the ZAMO's worldline is
${\p \over \p\tau} -y_o{\p \over \p\varphi}$.  If this were used to
define the energy flux, the energy flux would be in the same direction 
as the group velocity.

Quantum mechanically, the full 
extreme Kerr metric radiates particles even though
its Hawking temperature is zero. The emitted particles all lie in
the superradiant regime $\t \sigma < m \Omega _H$ \page.  
Since the modes we consider correspond to $\t \sigma = m \Omega _H$
one might wonder if this implies that the throat metric
will be quantum mechanically
stable. However the fact that the throat solution itself exhibits
superradiance suggests that it will decay quantum mechanically also.
Presumably, if one starts with an $SL(2,R)\times U(1)$ invariant vacuum
state, these symmetries will be preserved. It is then far from clear
what the solution \glob\ could decay into. This question requires further
investigation.

    Classically, the question of stability to linearized perturbations
is the question of whether there are unstable quasinormal modes obeying
outgoing wave boundary conditions at both infinities.  Our expectation,
in the absence of detailed investigation, is that no such
unstable quasinormal modes exist, based on the known stability of
Kerr black holes \det.  On the other hand, when back reaction is taken
into account the throat geometry is unstable, and nearby solutions
are singular. This follows from the singularity theorems \sing\ since
the constant $(y,\tau)$ two-spheres are marginally trapped. A generic
perturbation
will cause the orthogonal null geodesics to start converging, creating 
trapped surfaces and geodesic incompleteness. If the nearby solutions are
all black holes, then this instability need not be a serious problem. But
since the singularity theorems do not prove the existence of event horizons,
one does not yet know if more serious singularities can arise.
In order to apply this instability argument, one only needs to ensure that
the nearby spacetimes satisfy Einstein's equation with matter obeying
the weak energy condition.
The same argument applies to $AdS_2\times S^2$ and shows that this spacetime
is also unstable. (For another argument to this effect see \mms.)
However, this argument does not apply to products of higher dimensional $AdS$
and spheres, since the null geodesics orthogonal to the sphere consist of
an entire light cone in $AdS$ which is expanding. So in the higher dimensional
case, the spheres
are not marginally trapped and e.g. $AdS_5 \times S^5$ is stable.

The instability of the throat geometry may not be a serious obstacle to
constructing a dual quantum mechanical
 theory since it has been argued in the case of
$AdS_2\times S^2$ that the dual theory may
describe just the ``ground states" of string theory with these boundary 
conditions \mms.
We will see in section five that a similar argument applies to our throat
geometry.

\newsec{Extension to extreme Kerr-Newman throats}

While our main interest is in vacuum solutions, we note that our
results can easily be generalized to include the entire range
of extreme Kerr-Newman solutions describing rotating and charged black holes.
Their near horizon geometries smoothly interpolate between the solution 
\glob\ and $AdS_2\times S^2$. The Kerr-Newman
metrics have exactly the same form as \kerr, except that
$\Delta = {\t r}^2 -2 M r +a^2 +q^2$, where $q$ is the electric
charge and $a M$ is still the angular momentum; and the
$2 M \t r$ factor in $ \omega$ is replaced by
$\t r^2 +a^2 -\Delta$.  The extremal limit
corresponds to $M^2 = a^2 +q^2$, and the horizon is
at $\t r = M$  with area $4\pi(M^2 +a^2)$. If we define $r_0^2 \equiv M^2 +a^2$,
the throat metric is similar to \throat\ with only two modifications:
The factor $(1 +\cos^2\theta)/2$
becomes $1 -{a^2 \over r_0^2} \sin^2\theta$, and there is
a different coefficient in the frame-dragging angular
velocity.  Expanding $ \omega$ to first order in $\t r -M$,
we now obtain
\eqn\knom{ \omega = {a \over r_0^2} -{2 a M \over r_0^4}
(\t r -M) ,}
so the coefficient of $r d t$ in \throat\ is
$2 a M / r_0^4$ instead of $1 / r_0^2$. The near horizon limit
is now
$$ ds^2 = \(1 -{a^2 \over r_0^2} \sin^2\theta\) \[ -{r^2\over r_0^2} dt^2 +
  {r_0^2 \over  r^2} dr^2 + r_0^2 d\theta^2\] $$ 
\eqn\knthroat{+  r_0^2 \sin^2\theta \(1 -{a^2 \over r_0^2} \sin^2\theta\)^{-1} 
	\( d\phi + {2arM\over r_0^4} dt\)^2 }
Notice that when $a=0$, this metric reduces to $AdS_2\times S^2$ as expected.
The change in the  coefficient of the off diagonal term
 carries over to the change of the
coefficient in front of the log in the relation \phitran\
between $\phi$ and $\varphi$.  Otherwise, the transformation
to the global $AdS_2$ coordinates remains the same, and
the above changes carry over to the metric of the throat
geometry \glob\ in the global coordinates.

    An interesting point along the Kerr-Newman sequence is
when confinement of modes of scalar waves starts to
break down.  In the angular harmonic equation ${1 \over 4}
m^2 \sin^2\theta$ is replaced by ${a^4 \over r_0^4}
m^2 \sin^2\theta$, and a rough estimate of the smallest
separation constant for a given $m$ is
\eqn\kmin{ K_{min} = |m|(|m|+1) +
0.8{a^4 \over r_0^4}m^2.}
The $m^2$ term in the radial
equation is replaced by ${2 a^2 \over r_0^2}m^2$,
so confinement is broken when
\eqn\knconf{{4 a^2 M^2 \over r_0^4} +{2 a^2 \over r_0^2}
> {K_{min} \over m^2}.}
This happens first for $m^2 \gg 1$, when ${a^2 \over M^2}
\approx 0.242$.  The critical points for other types of
fields (e.g., electromagnetic or gravitational perturbations)
will differ somewhat from this.
~
\newsec{Searching for a Holographic Dual}

In light of the growing evidence in favor of the AdS/CFT correspondence,
it is natural to speculate that string theory on spacetimes that
approach \glob\ will have a 
dual holographic description. Since the boundary at infinity is effectively
one dimensional,
and $SL(2,R)$ is the conformal group of a line, one expects the dual
theory to be
a conformal quantum mechanical system.
We have already pointed out several ways in which  \glob\ 
  is qualitatively similar to $AdS_2\times
S^2$, which is the near horizon geometry of an extremely charged
four dimensional black hole.
So one might expect that the 
dual theories might be similar,
with the $SU(2)$ symmetry of $S^2$ broken to $U(1)$ in the vacuum case, 
breaking supersymmetry as well.
Unfortunately, $AdS_2\times S^2$ is currently the least well understood example
of the AdS/CFT correspondence. We currently have little information about
the structure of this quantum mechanical system.

Let us first try to  follow the procedure used to discover the original
AdS/CFT duality. Starting with the extreme black hole, we can decrease the
string coupling $g$ to obtain a weakly coupled string description of the
black hole states. The fact that the entropy of an extreme Kerr black hole
is independent of Newton's constant and simply given by the angular
momentum (which we still assume is positive), $S = 2\pi J$,
 strongly suggests that it has a simple microscopic
description. In fact, it may ultimately be simpler than the Reissner-Nordstrom
solution which requires four different charges from the fundamental string
standpoint. 
 Since an extreme Kerr black hole has no Ramond-Ramond
fields present, the states at weak coupling must be ordinary excited states
of the string. Unfortunately, the lack of supersymmetry makes it difficult
at present to give a precise identification of the states\foot{When charges
are present, one can give a precise counting of the states of certain extreme
charged and rotating black holes, even when they are far from the
supersymmetric state \hlm.}.
One can instead
 use the correspondence principle \hopo\ to roughly describe these
states as follows. 
Since angular momentum is quantized, we want to keep $J$ fixed as
we slowly decrease the string coupling. The mass of the black hole
is $G M^2 = J$  and its horizon radius is $r_+ = G M$, where
$G$ is the four dimensional Newton's constant. The correspondence
principle says that the black hole makes a transition to an excited string
state when its horizon size is of order the  string scale. Setting the
mass of the black hole equal to the mass of an excited string state at this
point yields
\eqn\mass{ M_{bh} \sim {\ell_s\over G} \sim {\sqrt N\over \ell_s} \sim M_s}
where $N$ is the string level.
Since $G \sim g^2 \ell_s^2$, this implies that $g \sim N^{-1/4}$ at the
transition point.
The black hole entropy is then 
\eqn\entropy{S_{bh} \sim G M^2 \sim M r_+ \sim M \ell_s \sim \sqrt N} 
Since $G M^2 = J$ and $J$ is fixed, the appropriate weakly coupled
string states are states with $N\sim J^2$ and angular momentum $J$.
(This was also noticed in \rusu.)
The number of such states is approximately $e^{\sqrt N}$ which agrees
with the entropy of the black hole.

States with $N\sim J^2$ and angular momentum $J$
are far from extremal string states. The minimum mass
string state carrying angular momentum $J$  has $N= J$, but there is only
 one such state, so it could never reproduce the entropy of the black hole.
One might ask what happens if one starts with a string state with less
energy than the states corresponding to the extreme black hole and increase
the coupling. It appears that you might be in danger of forming a naked
singularity. But, of course, this is not what happens. If $N<J^2$,
 then $M\ell_s < J$, but the minimum size of the string is $r_{\rm min} = J/M$,
so $r_{\rm min} > \ell_s$. This means that
the string cannot form a black hole at the string scale but 
only at a larger scale corresponding to a larger value of the string coupling.
Setting the masses equal at this transition point, the resulting black hole
will have a Schwarzschild radius at least $r_+ = J/M = a$ which would again
correspond to an extreme Kerr black hole.

In terms of trying to construct a dual theory, the obvious 
problem is that these excited string states are not stable. It is
not clear how to take 
an appropriate limit to decouple the bulk string states  and
 extract the dynamics of the states at  a given level $N\sim J^2$.

A clue to the correct description may be the following:
Consider the form of the solution \knthroat. The area of the horizon
at $r=0$ is $A=4\pi r_0^2$ where $r_0^2 = 2a^2 +q^2$.
 Thus the extreme black hole entropy is related to 
the radius of curvature of the
\A\ space along the axis 
in a universal way which is independent of the ratio of charge to
angular momentum.
This suggests that there may be a unified description of string theory
with these boundary conditions.

It was pointed out in \mms\ that the energy above extremality $E$
of a near extremal
Reissner-Nordstrom black hole scales with temperature like 
$E=2\pi^2 Q^3 T^2 \ell_p$ where $\ell_p$ is the Planck length. The fact
that $\ell_p$ enters this formula means that one cannot take $\ell_p \a 0$
keeping $E, Q, T$ fixed. It is easy to see that the same thing is true
for the entire family of
near extreme Kerr-Newman solutions. For simplicity, we restrict here to 
the pure Kerr case. Including factors of Newton's
constant, the Hawking temperature is 
\eqn\hawktemp{T={(G^2M^2 -a^2)^{1/2} \over 4\pi GM[GM + (G^2M^2 -a^2)^{1/2}]}}
In the extremal limit, $G M =a$. Let $GM = a + \epsilon$, and define
$E=M - (a/G)= \epsilon/G$. Then we have 
$T = (2GM\epsilon)^{1/2}/4\pi G^2 M^2$, so $E = 8\pi^2 T^2 G^2 M^3$.
But $GM^2 =J$ so 
\eqn\temp{ E = 8\pi^2  J^{3/2} T^2 \ell_p }

\newsec{Five Dimensional Throat Geometries}

Generalizations of the Kerr metric to more than three space dimensions have been
discussed thoroughly by Myers and Perry \mype. 
 In $N$ space (plus one time) dimensions
there are $[N/2]$ independent planes of rotation in which axial symmetry can
be enforced simultaneously, with the same number of independent angular momentum
parameters.  The spatial coordinates reflecting this symmetry consist of
$[N/2]$ axial angles, $[(N-1)/2]$ polar angles, and one radial coordinate.
Extremal vacuum black holes with a non-singular degenerate horizon do not exist
in all cases, but when they do, we expect that the near horizon
limit leads to a complete spacetime
with an $SL(2,R) \times U(1)^{[N/2]}$ symmetry group.  As an
illustrative example, we work through some of the details for the 5D ($N = 4$)
case.

In five dimensions, let $\t \phi$ and $\t \psi$ denote the two axial angles,
$a$ and $b$ the corresponding angular momentum parameters, and define
a mass parameter $\mu$ with units of length squared.  The physical mass of
the black hole is $M = 3 \pi \mu/ 8 G_5$, and the factor $2M/3$
converts angular momentum parameters to physical angular momenta.
The metric in ``Boyer-Lindquist" coordinates (Equation (3.18) of \mype)
reduces to
$$ds^2 = -d{\t t}^2 + {\mu \over \rho^2}(d \t t -a \sin^2\theta d \t \phi
-b \cos^2\theta d \t \psi)^2$$
\eqn\kerrfive{+ (\t r^2 +a^2)\sin^2\theta d \t \phi^2
+ (\t r^2 +b^2)\cos^2\theta d \t \psi^2 + \rho^2 d \theta^2
+ {\rho^2 \t r^2 \over \Delta} d \t r^2,}
where
\eqn\defsfive{ \rho^2 \equiv \t r^2 +a^2 \cos^2\theta +b^2 \sin^2\theta,
\qquad \Delta \equiv (\t r^2 +a^2)(\t r^2 +b^2) - \mu \t r^2.}
The range of the polar angle $\theta$ is $0 \le \theta \le \pi/2$.
Since no odd powers of $\t r$ are present, it is convenient to
work with $\t u \equiv \t r^2$ as the radial coordinate, with
\eqn\guu{g_{\t u \t u} = {\rho^2 \over 4 \Delta}.}

The horizon is at the outermost zero of $\Delta$.  This is a double
zero, implying an extremal horizon, if $\mu = (|a| +|b|)^2$.  The
extremal horizon is at $\t u = |a b|$.  Unless both angular momentum
parameters are non-zero the horizon
is singular, since $\rho^2$ on the horizon is zero at
$\theta = 0$ if $a = 0$ or at $\theta = \pi/2$ if $b = 0$, and
$\rho^2 = 0$ implies a curvature singularity.  The 3-volume of
the horizon is $2 \pi^2 |a b|^{1/2} (|a| +|b|)^2$.

We focus on the inverse metric tensor,
since it is the inverse metric tensor that comes into the Carter
Hamilton-Jacobi formulation of the geodesic equations and the
scalar field equation.  In particular, note that the coordinate
angular velocities of a ZAMO are given by
$\omega^{\t \phi} = g^{\t \phi \t t}/g^{\t t \t t}$
and $\omega^{\t \psi} = g^{\t \psi \t t}/g^{\t t \t t}$,
or
\eqn\omphps{\omega^{\t \phi} = {a \mu (\t u +b^2) \over \Sigma},
\qquad \omega^{\t \psi} = {b \mu (\t u +a^2) \over \Sigma},}
where
\eqn\sigmadef{\Sigma \equiv \mu (\t u +a^2)(\t u +b^2) +\Delta \rho^2.}
The inverse square of the lapse function is
\eqn\invlapse{g^{\t t \t t} =- {\Sigma \over \Delta \rho^2}.}
The axial components of the inverse metric tensor are
\eqn\invphph{g^{\t \phi \t \phi} = {g_{\t \psi \t \psi} \rho^2 \over
\Sigma \sin^2\theta \cos^2\theta} + ( \omega^{\t \phi})^2 g^{\t t \t t},}
\eqn\invpsps{g^{\t \psi \t \psi} = {g_{\t \phi \t \phi} \rho^2 \over
\Sigma \sin^2\theta \cos^2\theta} + ( \omega^{\t \psi})^2 g^{\t t \t t},}
\eqn\invphps{g^{\t \phi \t \psi} = -{g_{\t \phi \t \psi} \rho^2 \over
\Sigma \sin^2\theta \cos^2\theta} +  \omega^{\t \phi}
 \omega^{\t \psi} g^{\t t \t t}.}
Also, we note that
\eqn\sqg{(-g)^{1/2} = {1 \over 2} \rho^2 \sin\theta \cos\theta}
using $\t u$ as the radial coordinate.

The throat limit is obtained by changing the axial coordinates to be
corotating with the horizon and then rescaling the $\t u$ and $\t t$
coordinates, similar to what was done in \limit.  Specifically,
\eqn\limfive{\t \phi = \phi + {a \over |a b| +a^2} \t t, \quad
\t \psi = \psi + {b \over |a b| +b^2} \t t, \quad \t u = |a b| +
\lambda (|a|+|b|)^2 u, \quad \t t = {\sqrt{|ab|} \over 2\lambda} t.}
In the limit $\lambda \a 0$, $\rho$ becomes a function of
$\theta$ only, $\rho^2 = |ab| + a^2\cos^2\theta + b^2 \sin^2 \theta$,
and the components of the inverse metric become
\eqn\ilaplim{g^{t t} = -{4 \over  \rho^2 u^2},
\qquad g^{u u} = {4 u^2 \over \rho^2}, 
\qquad g^{\theta \theta} = {1 \over \rho^2},}
\eqn\omphpslim{\omega^\phi = g^{\phi t}/g^{t t} =
-{u\over 2}{a\over |a|}\bigg|{b\over a}\bigg|^{1/2},
 \qquad \omega^\psi = g^{\psi t}/g^{t t}
= -{u\over 2}{b\over |b|}\bigg|{a\over b}\bigg|^{1/2},}
\eqn\phphlim{g^{\phi \phi} = {|b| +|a| \cos^2\theta \over
|a|(|a| +|b|)^2 \sin^2\theta} +(\omega^\phi)^2 g^{t t},}
\eqn\pspslim{g^{\psi \psi} = {|a| +|b| \sin^2\theta \over
|b|(|a| +|b|)^2 \cos^2\theta} +(\omega^\psi)^2 g^{t t},}
\eqn\phpslim{g^{\phi \psi} = -{a b \over |a b|(|a| +|b|)^2}
+ \omega^\phi \omega^\psi g^{t t},}

Despite its rather complicated appearance, one immediately sees the
characteristic structure of \A\ in \ilaplim, with the horizon at $u=0$.
Eqs. \phphlim\ - \phpslim\ are all just functions of $\theta$.
This metric has essentially the same set of $AdS_2$-like Killing
vector fields as the 4-D Kerr throat metric.  An almost identical
transformation to global time coordinate $\tau$, radial
coordinate $y$, in which
\eqn\newfive{u =  \[{(1 +y^2)^{1/2} \cos\tau +y}\], \qquad
t = { (1 +y^2)^{1/2} \sin\tau \over u}}
and new axial angle coordinates $\varphi$ and $\chi$
related to $\phi$ and $\psi$ by expressions like \phitran\
with appropriate coefficients in front of the logs, gives a globally
non-singular metric describing a geodesically complete spacetime.
The total number of Killing fields is five, since there are two axial Killing
fields
instead of one.

Two special cases are worth mentioning.
When the two angular momenta components are equal,
the metric simplifies and $\rho^2 = 2a^2$ no longer depends on $\theta$.
 Setting $a=b>0$, the metric
 (not the inverse metric)
becomes
$$ ds^2 = -{a^2\over 2} u^2 dt^2 + {a^2\over 2} {du^2 \over u^2} 
+2a^2 d\theta^2   $$ 
$$+2a^2\[\sin^2\theta\(d\phi + {u\over 2} dt\)^2
           +\cos^2\theta\(d\psi + {u\over 2} dt\)^2\]$$
\eqn\eqang{ +2a^2\[\sin^2\theta \(d\phi + {u\over 2} dt\) + \cos^2 \theta
         \(d\psi + {u\over 2} dt\) \]^2 }
Along a zero angular momentum observer (ZAMO) worldline,
 $d\phi + {u\over 2} dt =0$ and 
$d\psi + {u\over 2} dt=0$.

When one of the angular momenta vanishes, the horizon of the extreme
black hole becomes singular. It is a surprising fact that if one takes
the near horizon limit in this case, one finds a spacetime with $AdS_3$
symmetry\foot{We thank Kirill Krasnov for pointing this out to us.}!
We start with the general solution \kerrfive\ with $b=0$, and set
$\mu=a^2$ to obtain the extremal limit. We then shift  
$\t \phi = \phi + \t t/a$, and rescale $ \t r = \lambda r, \t t= t/\lambda, \
 \t\psi = \psi/\lambda$ taking $\lambda \a 0$. The result is
\eqn\adsthree{ds^2 = \cos^2\theta\[-{r^2\over a^2} dt^2 + {a^2\over r^2} dr^2
  + r^2 d\psi^2\] + a^2 \[\cos^2 d\theta^2 + {\sin^2\theta\over \cos^2\theta}
   d\phi^2\] }
Strictly speaking, $\psi$ should have only infinitesimal extent, since
$\t \psi$ is periodic and $\psi = \lambda \t \psi$. However, we can clearly
extend $-\infty < \psi < \infty$ and obtain
a five dimensional vacuum solution which has the $SO(2,2)$ 
symmetries of $AdS_3$. Recall that $0\le \theta \le \pi/2$. The above 
solution looks like a product of $AdS_3$ and a disk, but is singular
at $\theta = \pi/2$.  It is not yet clear whether this singularity
justifies throwing this solution away as unphysical. There are many 
examples of singular solutions which play a prominent role in string
theory (e.g. the metric of most D-branes). Yet some singular metrics are
clearly unphysical (e.g. the negative mass Schwarzschild solution).
The prominent role of $AdS_3$ in this vacuum solution justifies further
investigation.

Solutions of the geodesic equations in the general metric \ilaplim\ - \phpslim\
 can be obtained by separation
of variables in the Hamiltonian-Jacobi formalism \carter.  The
generating function $S$ for the canonical transformation to
coordinates which are constants of the motion obeys an equation
derived by substituting ${\p S \over \p x^\alpha}$ for $p_\alpha$
in the expression $g^{\alpha \beta}p_\alpha p_\beta = -\mu^2 =
{\p S \over \p \lambda}$, where $\lambda$ is the affine parameter.
Since $p_t = -E$, $p_\phi = L_\phi$, and $p_\psi = L_\psi$ are
trivial conserved quantities, the equation for $S$ becomes
$$4 u^2 \({\p S \over \p u}\)^2-{4 \over u^2}
 \[{E + { u \over 2}{a\over |a|}\bigg|{b\over a}\bigg|^{1/2}
L_\phi + { u \over 2}{b\over |b|} \bigg|{a\over b}\bigg|^{1/2} L_\psi}\]^2$$
$$ - {2 a^2 -b^2 \over
|a| (|a| +|b|)}L_\phi^2 - {2 b^2 -a^2 \over |b| (|a| +|b|)}L_\psi^2
- {a b \over |a b|}L_\phi L_\psi + |a b| \mu^2$$
$$+ \({\p S \over \p \theta}\)^2 
+{L_\phi^2 \over \sin^2\theta} +{|a| -|b| \over |a| +|b|}
L_\phi \({L_\phi + {a b \over |a b|}L_\psi}\) \sin^2\theta
+ b^2 \mu^2 \sin^2\theta $$
\eqn\hamjac{+{L_\psi^2 \over \cos^2\theta} +{|b| -|a| \over |a| +|b|}
 L_\psi \({L_\psi + {a b \over |a b|}L_\phi}\) \cos^2\theta
+ a^2 \mu^2 \cos^2\theta,} 
and
\eqn\genfn{S = -\mu^2 \lambda - E t + L_\phi \phi + L_\psi \psi
+ \int{ \Theta(\theta)^{1/2} d \theta}
+ \int{R(u)^{1/2} d u}.} 
 Using a constant $K$ to separate
the $u$-dependence of the first two lines of \hamjac\ from the
$\theta$-dependence of the third and fourth lines gives
$$\Theta(\theta) = K -{L_\phi^2 \over \sin^2\theta} -{|a| -|b|
\over |a| +|b|} L_\phi \({L_\phi + {a b \over |a b|}L_\psi}\)
\sin^2\theta -  b^2 \mu^2 \sin^2\theta $$
\eqn\anggen{-{L_\psi^2 \over \cos^2\theta}
-{|b| -|a| \over |a| +|b|} L_\psi \(L_\psi + {a b \over
|a b|}L_\phi\) \cos^2\theta - a^2 \mu^2 \cos^2\theta}
and
$$ 4 u^2 R(u) = {4 \over u^2} \[{E + {a u \over 2|a|}
\bigg|{b\over a}\bigg|^{1/2} L_\phi + {b u \over 2|b|} 
\bigg|{a\over b}\bigg|^{1/2} L_\psi}\]^2 $$
\eqn\radgen{+ {2 a^2 -b^2 \over |a| (|a| +|b|)}L_\phi^2 + {2 b^2 -a^2 \over
|b| (|a| +|b|)}L_\psi^2 + {a b \over |a b|}L_\phi L_\psi - |a b| \mu^2
- K.}

Geodesics are obtained by setting the partial derivatives of $S$ with respect
to $\mu,E,L_\phi,L_\psi,K$ equal to constants (which reflect initial conditions
for the geodesics). This directly gives $\lambda, t,\phi,\psi$ as functions
of $\theta$ and $u$ along the geodesic, as well as the relation between
$\theta$ and $u$. 
The range of radial motion of the test particle is where $R(u) > 0$.
This extends to infinity if
\eqn\confive{{2|a| +|b| \over |a| +|b|}L_\phi^2 + {3 a b \over |a b|}
L_\phi L_\psi + {2|b| +|a| \over |a| +|b|}L_\psi^2 > K + |a b| \mu^2.}
From \anggen\ a trajectory at $\theta = 0$ has $L_\phi = 0$
and $K = {2|b| \over |a| +|b|}L_\psi^2 +a^2 \mu^2$.  A trajectory
 at $\theta = \pi/2$ has $L_\psi = 0$ and $K = {2 |a|
\over |a| +|b|} L_\phi^2 + b^2 \mu^2$.  In both cases there are
radially unbounded trajectories if the appropriate angular momentum is
sufficiently large.  The extension to modes of scalar waves
is straightforward, and again there will be some modes with non-zero
axial eigenvalues $m_\phi$ and/or $m_\psi$ which propagate as
oscillating waves to infinity. Superradiant scattering will exist, though
the condition for superradiance is more complicated.
 The qualitative picture is just
as it was in the four dimensional case.

\newsec{Conclusion}

We have explored the near horizon geometry of an extreme rotating
black hole in the hopes that it will be useful in extending the
remarkable duality between string theory and field theory to the
vacuum case. It is not yet clear whether this will succeed. While
the vacuum solutions we have discussed have striking similarities
to the $AdS_2\times S^2$ geometry arising in the near horizon limit of
extreme Reissner-Nordstrom black holes, there are some crucial differences. 
These include the fact that in addition to the usual localized modes of
a test field with discrete frequencies, there are also traveling waves
with continuous frequencies that exhibit a type of superradiance.
The quantum analog of this superradiance, and its possible implications
for duality, remains to be explored.

Another open question raised by this work is the nature of the singularity
in the five dimensional vacuum solution with $AdS_3$ symmetry discussed
in the previous section. Does this solution have physical interest?

\centerline{\bf Acknowledgements}

\vskip .2cm
It is a pleasure to thank K. Krasnov and R. Myers for discussions. 
This work was supported
in part by NSF Grants PHY94-07194 and PHY95-07065.

\listrefs
\end